\documentclass[doublecol,linenumbers]{epl2}
\usepackage{amsmath}
\usepackage{amssymb}
\usepackage{amsfonts}
\usepackage{graphicx}
\usepackage{dcolumn}
\usepackage{bm}
\usepackage{sidecap}
\usepackage {caption}
\usepackage{float}
\usepackage{parskip}
\usepackage{array,booktabs,arydshln,xcolor}

\usepackage{hyperref}

\usepackage{hyperref}
\usepackage{epstopdf}
\makeatletter
\def\@eqnnum{{\normalsize \normalcolor (\theequation)}}
 \makeatother

\title{Interplay of inhibition and multiplexing : Largest eigenvalue statistics}

\author{Saptarshi Ghosh{$^1$}, Sanjiv K. Dwivedi{$^1$}, Mikhail V. Ivanchenko{$^2$}
and Sarika Jalan{$^1$}\footnote{Corresponding Author:sarikajalan9@gmail.com}}

\institute{1. Complex Systems Lab, Discipline of Physics, 
Indian Institute of Technology Indore, Simrol, Indore-453552, India\\
2.
Department of Applied Mathematics, Lobachevsky State University of Nizhny Novgorod, Russia}


\abstract{The largest eigenvalue of a network provides understanding to various dynamical as well as 
{stability} properties of the underlying system. We investigate interplay of inhibition and multiplexing on the 
largest eigenvalue {statistics} of networks. Using numerical experiments, we demonstrate that 
presence of the inhibitory coupling may lead to a very different behaviour of the largest eigenvalue statistics of 
multiplex networks than those of the isolated networks depending upon network architecture of the 
individual layer. We demonstrate that there is a transition from the Weibull to the Gumbel or to the Fr\'echet distribution as networks are multiplexed. Furthermore, for denser networks, there is a convergence to the Gumbel distribution as network size increases indicating higher stability of larger systems.}
\begin{document}

\maketitle

{\it Introduction:} In recent years, network science has attracted researchers from diverse communities owing to its remarkable applicability to understand behavior of {many} real world complex systems \cite{barabasi_review}. One of the widely investigated areas in the network science {is understanding} relation of spectral properties of {the network} adjacency matrices with {dynamical and structural properties of corresponding systems.} {Particularly}, the largest eigenvalue of a network adjacency {or coupling matrix} has been {shown} to strongly influence dynamical evolution on the corresponding network, {for instance, synchronization properties of coupled Kuramoto oscillators \cite{sync_osc} and dynamical properties of neural network \cite{neural}}. Further, {stability of ecological systems are demonstrated to be related with the largest eigenvalue of the interaction matrix of different species \cite{may_stability}}. 

Furthermore, it has been increasingly realized that multiplex networks provide a better framework to 
{investigate structural and dynamical} properties of many real world complex systems \cite{mul_org_1}. 
{As defined in \cite{delay_multiplex}, a multiplex network consists of different layers with one to one 
correlation between the mirror nodes of different layers}. Few examples of complex systems, which
can be represented in multiplex network framework, are banks, transportation, stock market, etc \cite{mul_exp}. 
A node in the multiplex network architecture may have a very different dynamical behavior due to the 
influence of other layers than it has in a single layer network \cite{delay_multiplex}. 
{Further}, inhibition in the coupling is known to play a crucial role in functioning and evolution of many 
real-world networks including brain and ecological networks \cite{inhibition}. 
We introduce inhibitory and excitatory coupling in different layers of the 
multiplex networks and investigate statistical properties of $R_{\mathrm max}$.
In this Letter, we analyze interplay of multiplexing and inhibition on behavior of 
{$R_{\mathrm max}$ of an ensemble of networks coupling matrices 
under GEV framework.} The GEV statistics of independent, identically distributed random variable has been 
successfully applied to many real world systems including stock markets, natural disasters, 
galaxy distributions as a model for extreme events \cite{gev_book}. 
	
{One such system having multiplex network architecture as well as inhibition in the coupling is 
ecological system, which for instance, may consist of different geographical regions represented as 
different layers. These layers possess inhibitory and excitatory coupling
due to presence of predator-prey, mutualistic and competetative relations among the species. 
Further, species at different geographical regions may interact with each other due to, for instance, 
migration. $R_{\mathrm max}$ statistics indicates stability of such ecosystems, for example, 
against extinction of a species}. 
We investigate impact of {inhibition} on $R_{\mathrm max}$ behaviour {of multiplex networks and compare
them with those of the isolated networks}. Since, different layers in a multiplex network can have different 
network architectures, we explore impact of mixing of different types of network architectures on 
$R_{\mathrm max}$ statistics.

\begin{figure}[t]
 \centerline{\includegraphics[width=1.7in, height=1.3in]{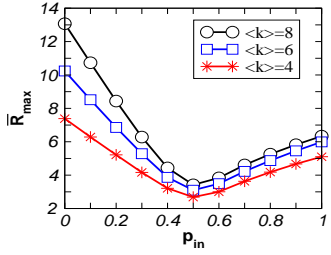}}
 \caption{(Color online) Average ${R}_{\mathrm max}$ as a function of inhibition probability ($p_{in}$) for 
SF networks of various average degrees { $\langle k \rangle$}. Size of the network 
remains N=100. Average is taken over $5000$ random realizations of the network.} 
 \label{r_max_avg}
 \end{figure}

{\it Theoretical framework:} The adjacency matrix $A$ of a network has entries $A_{ij}=1$ or $0$ depending 
upon whether $i$ and $j$ nodes are connected or not. The diagonal entries of $A$ are zero depicting 
no self connection. We use the Erd\"{o}s-R\'enyi (ER) model to generate {a} 
random network \cite{barabasi_review}. {Further,} we use configuration model \cite{config} to generate 
scale-free (SF) networks.  To begin with, we consider a network comprising all the excitatory connections leading 
to a symmetric adjacency matrix {i.e. $A_{ij}=A_{ji}=1$} having {only} $0$ and $1$ matrix elements. 
An introduction of the {inhibitory nodes,} with probability $p_{in}$, {replaces all $1$ entires to} 
$-1$ in the corresponding rows of the adjacency matrix 
and consequently symmetric property of the matrix is lost \cite{pin_prob} {as a row corresponding to a inhibitory
node will have all $-1$ non-zero entires, whereas corresponding column  may consist of elements having different signs
depending upon if they belong to the inhibitoy or excitatory nodes.
We henceforth declare the adjacency matrix with the directional signs as coupling matrix \cite{cou_mat_PRE}}
and investigate distribution of (${R}_{\mathrm max}$) of the 
ensemble of these matrices having $-1,0,1$ entries. 

Further, for the multiplex network, {coupling} matrix $A$ can be defined as, 
\begin{equation}
   A=
      \begin{pmatrix} A^{(1)} & I \\ I & A^{(2)} \end{pmatrix}, 
\label{mul_mat}
\end{equation}
where $A^{(1)}$ ($A^{(2)}$) represents {coupling} matrix of the first (second) layer and $I$ is an
unit $N$X$N$ matrix where $N$ is the size of  $A^{(1)}$ and $A^{(2)}$ networks. We start with $P$ random realizations of the multiplex network with ER-ER, SF-SF and ER-SF network topologies and introduce inhibitory couplings in each layer with probability {$p_{in}^{(1)}$} and {$p_{in}^{(2)}$}, respectively and {investigate the effect of inhibition in individual layers on the collective $R_{\mathrm max}$ behavior of the multiplex network.}

GEV distributions can be characterized entirely in terms of three universal probability distribution functions (PDF) namely Weibull, Gumbel and Fr\'echet depending on the tail of density function being power law, faster than power law and bounded or unbounded, respectively \cite{gev_book}. The probability density function for three GEV distributions can be written as
 	{ \small
 		\begin{equation}
 		\rho(x) = \begin{cases} \frac{1}{\sigma}\big[1+\big(\xi\frac{(x-\mu)}{\sigma}\big)\big]^{-1-\frac{1}{\xi}} \exp\big[-\big(1+\big(\xi\frac{(x-\mu)}{\sigma}\big)\big)^{-\frac{1}{\xi}}\big]&\\
 		\hspace*{5.6cm}\mbox{if } \xi\not=0 \\
 		\frac{1}{\sigma}\exp\big(-\frac{x-\mu}{\sigma}\big)\exp\big[-\exp\big(-\frac{x-\mu}{\sigma}\big)\big]
 		~~ \mbox{if } \xi=0. \end{cases}
 		\label{eq_gev}
 		\end{equation}
 	}
where $\mu$, $\sigma$, $\xi$ represent location parameter, scale parameter and shape parameter, respectively. The underlying statistics can be determined by the value of the shape parameter as follows; $\xi > 0$ (Fr\'echet Statistics), $\xi = 0$ (Gumbel statistics) and $\xi < 0$ (Weibull statistics).  We use Kolmogorov-Smirnov (KS) test \cite{KS_test} to characterize the distributions in terms of GEV statistics derived from the numerical simulations. 
\begin{figure}[t]
\centerline{\includegraphics[width=3.0in, height=2.0in]{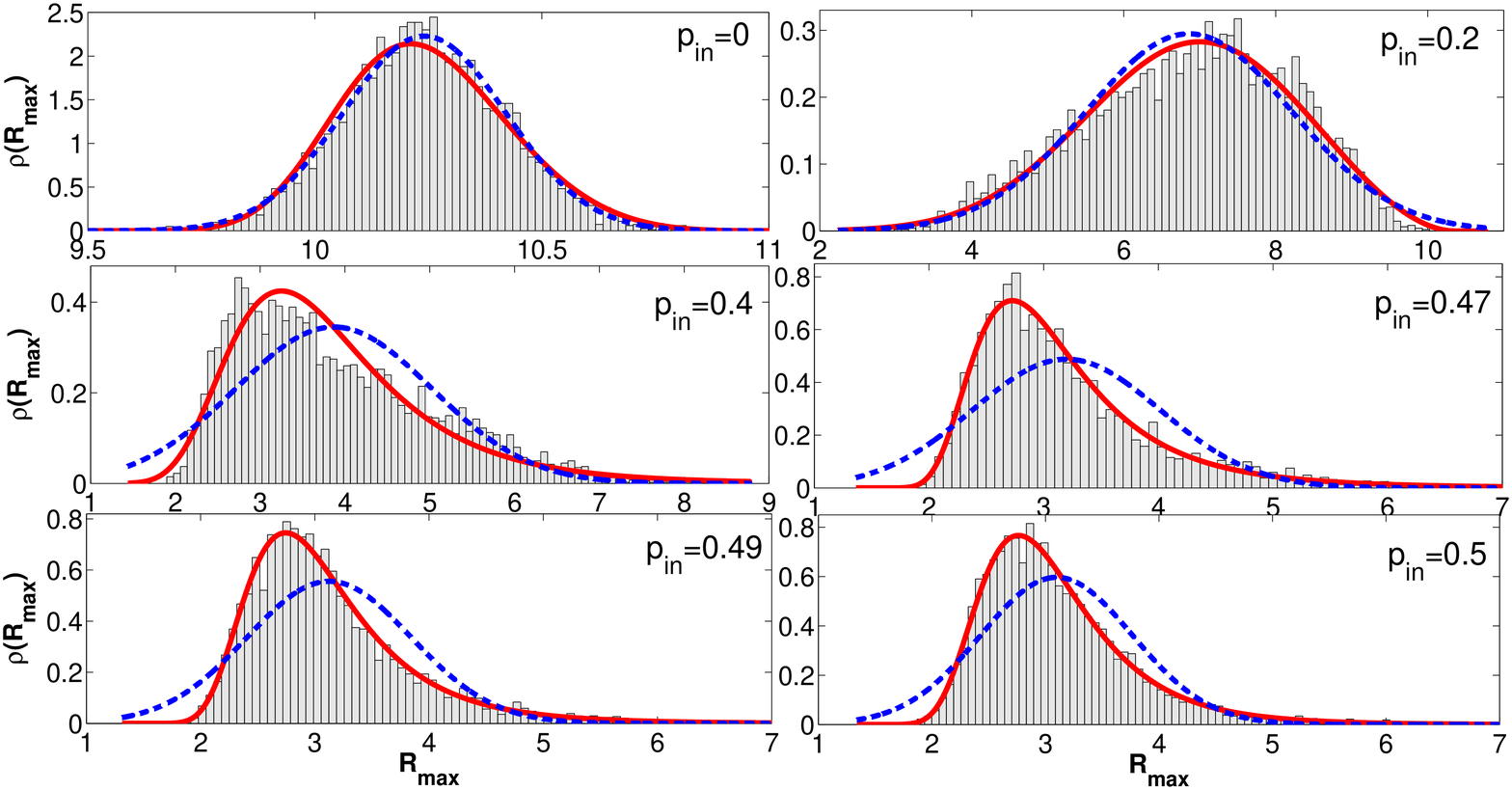}}
\caption{(Color online) Distribution of $R_{max}$ for SF networks with $\langle k\rangle=6$ and
for various inhibitory probabilities ($p_{in}$). 
Histogram is fitted with normal (blue dotted line) and GEV (red solid line) distributions.}
\label{his_sf_iso}
\end{figure}

{\it Largest eigenvalue statistics for isolated network:} We start the investigation for the 
isolated ER and SF networks followed by the multiplex network having different network topologies representing each layer.
First, we discuss the average behavior of $R_{\mathrm max}$ for the isolated SF networks. As inhibitory couplings are introduced in the network, thereby leading to asymmetricity in the {coupling} matrix, spectra of network might start taking up complex eigenvalues. Consequently, the average value of $R_{\mathrm max}$ decreases up to $p_{in}\le0.5$ (Fig.\ref{r_max_avg}).
At $p_{in} = 0.5$, $\overline{R}_{\mathrm max}$ shows the minimum value by displaying the global {minima}. The smallest eigenvalue of a network adjacency matrix ($\lambda_{min}$) with all the positive entries becomes the largest eigenvalue ($\lambda_{\mathrm max}$) of the same matrix with all the negative entries ($p_{in}=1$) and as $\lambda_{\mathrm max}\ne \lambda_{min}$ for the SF network with all the positive entries. Thus the behaviour of $R_{\mathrm max}$ becomes asymmetric about the minima at $p_{in}=0.5$ (Fig.\ref{r_max_avg}). A similar behavior is observed for different average degrees of the SF networks (Fig.~\ref{r_max_avg}).

\begin{figure}
\centerline{\includegraphics[width=3.0in, height=2.0in]{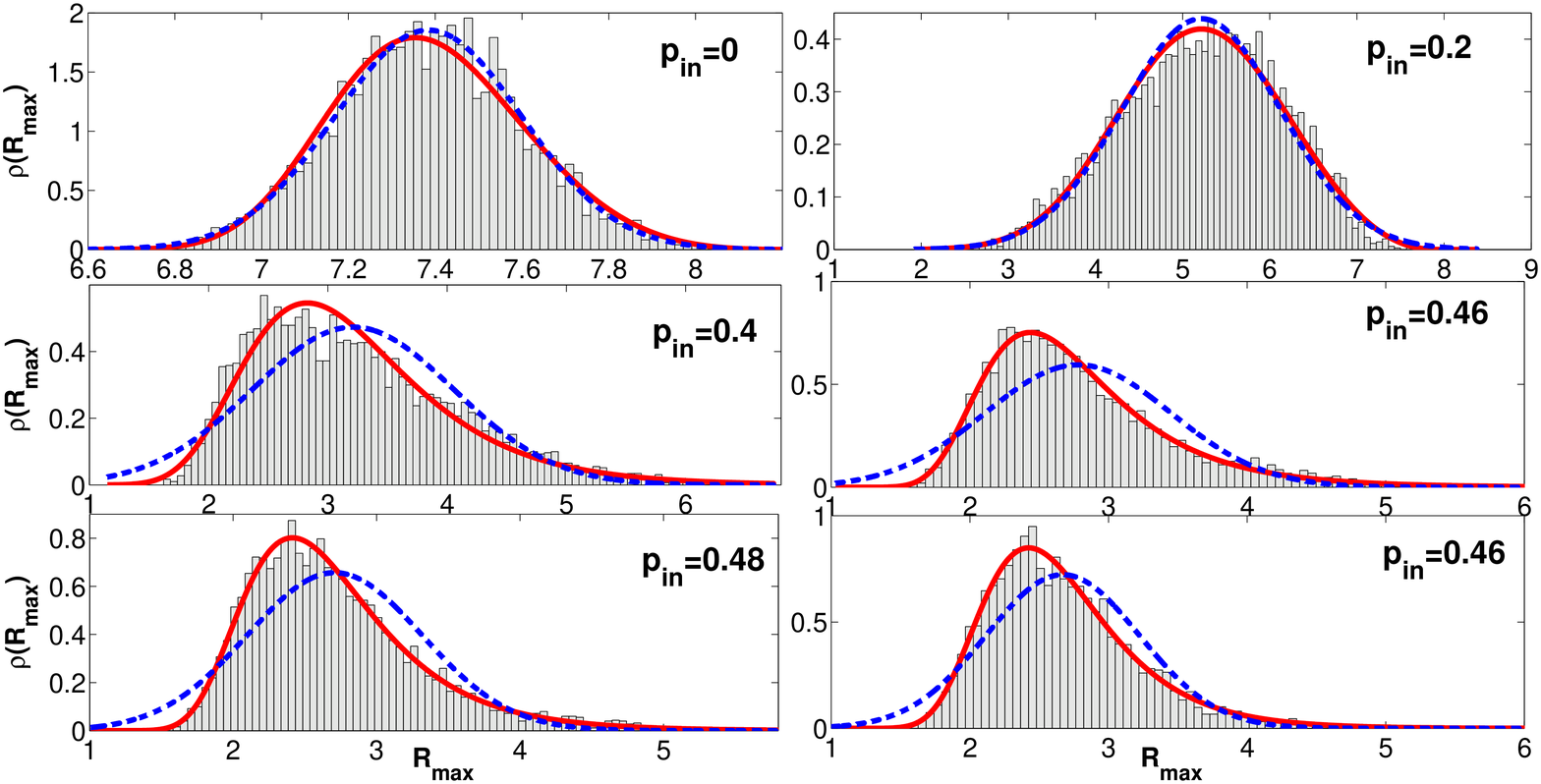}}
\caption{(Color online) Distribution of $R_{max}$ for 5000 population of SF networks with average degree $\langle k\rangle=4$ for various values of inhibitory probability ($p_{in}$). Histogram is fitted with normal (blue dotted line) and GEV (red solid line) distributions.}
\label{his_sf_iso4}
\end{figure}

Next, we probe for the fluctuations in the largest eigenvalue around the mean value. The statistics of $R_{\mathrm max}$ is fitted with the normal and GEV distribution (Eq.\ref{eq_gev}). For $p_{in}=0$, we find that the distribution can be modeled using the GEV statistics and value of the shape parameter characterizes the statistics as the Weibull distribution {(Figs.~\ref{his_sf_iso} and ~\ref{his_sf_iso4})}. Details of fitting parameters {for different degrees are given in Tables~\ref{table.1} and ~\ref{table.2}}. Note that the KS test accepts the normal distribution as well for $p_{in}=0$. This is due to a close resemblance of the Weibull distribution with the normal distribution for a particular parameter regime \cite{GEV_normal}. { 
Furthermore, $R_{min}$
lying in the left tail of the triangular shape distribution of eigenvalues of an ensemble of
SF networks (at $p_{in}=0$) is known to follow
a power law behaviour \cite{edge_behaviour} which is bounded due to the finite size effect. This
combination charecterizes $R_{min}$ for $p_{in}=0$ and consequently $R_{max}$ for $p_{in}=1$ as 
the Weibull distribution.} For the intermediate $p_{in}$ values, except at $p_{in}=0.5$, certain $p_{in}$ values can be modeled using the GEV statistics but without any consistent behavior. 

\begin{largetable}
\caption{Estimated parameters of KS test for fitting of GEV and normal distributions of $R_{\mathrm max}$ for different
inhibitory inclusion probability ($p_{\mathrm in}$) of SF network over 5000 population. Other parameters are network size $N = 100$ and average degree $\langle k\rangle$ = 6.} 
\begin{tabular} { c  c  c  c  c  c  c  c }     
\hline  \specialrule{0.5pt}{0pt}{0pt}
\scriptsize $p_{\mathrm in}$ & \scriptsize $\xi$ of GEV & \scriptsize $\sigma$ of GEV & \scriptsize $\mu$ of GEV  & \scriptsize p-value of KS test for GEV & \scriptsize $\mu$ of Normal & \scriptsize $\sigma$ of Normal & \scriptsize p-value of KS test for Normal\\ \hline  \specialrule{1pt}{0pt}{0pt}
\scriptsize 0.0& \scriptsize	-0.2065& \scriptsize	0.1758& \scriptsize	    10.1715& \scriptsize	0.0125& \scriptsize	    10.2413& \scriptsize	0.1790& \scriptsize	    0.1158	\\  
\scriptsize 0.2&	\scriptsize -0.3580&	\scriptsize 1.4003&		\scriptsize 6.4256&		\scriptsize 0.0191&		\scriptsize 6.8590&		\scriptsize 1.3539&		\scriptsize 0.0000	\\   
\scriptsize 0.4&	\scriptsize 0.0663&		\scriptsize 0.8681&		\scriptsize 3.3062&		\scriptsize 0.0000&		\scriptsize 3.8690&		\scriptsize 1.1556&		\scriptsize 0.0000	\\  
\scriptsize 0.47&	\scriptsize 0.1586&		\scriptsize 0.5245&		\scriptsize 2.8023&		\scriptsize 0.1285&		\scriptsize 3.1987&		\scriptsize 0.8179&		\scriptsize 0.0000	\\   
\scriptsize 0.49&	\scriptsize 0.0937&		\scriptsize 0.4955&		\scriptsize 2.7889&		\scriptsize 0.4406&		\scriptsize 3.1253&		\scriptsize 0.7177&		\scriptsize 0.0000	\\   
\scriptsize 0.50&	\scriptsize 0.0597&		\scriptsize 0.4807&		\scriptsize 2.7905&		\scriptsize 0.9688&		\scriptsize 3.0983&		\scriptsize 0.6684&		\scriptsize 0.0000	\\ 
\hline \specialrule{0.5pt}{0pt}{0pt}
\end{tabular}
\begin{flushleft} 
\label{table.1} 
\end{flushleft}
\end{largetable}

\begin{figure}
\centerline{\includegraphics[width=3.1in, height=1.4in]{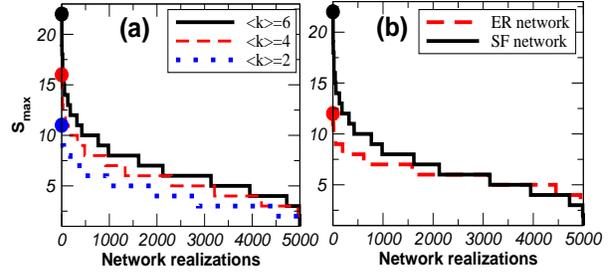}}
\caption{(Color online) $S_{\mathrm max}$ as a function of random realizations of the matrices. (a) $S_{\mathrm max}$ for SF networks with $\langle k \rangle=6$ (black solid line ), $\langle k \rangle=4$ (red dashed line), $\langle k \rangle=2$ (blue dotted line). (b) $S_{\mathrm max}$ for average degree $\langle k \rangle=6$ for SF (black solid line) and 
ER (red dashed line) networks. Other parameters are N=100 and $p_{in}=0.5$. The graph is plotted such that network with the highest $S_{\mathrm max}$ comes first.}
\label{cmax}
\end{figure}

At $p_{in}=0.5$, the statistics can be modeled using GEV statistics and exact form of the distribution depends on the average degree and the network size. At this $p_{in}$ value, a transition is observed from the Weibull to the Fr\'echet via the Gumbel distribution as denseness {($\langle k \rangle$)} of the network increases {(
Figs.~\ref{his_sf_iso} and ~\ref{his_sf_iso4})}.
For a small network size, the reason behind this transition can be explained in terms of $S_{\mathrm max}$ behaviour of the network. $S_{\mathrm max}$ represents the maximum value of the column sum of a particular {coupling} matrix in a network ensemble. Figure \ref{cmax} (a) displays distribution of $S_{\mathrm max}$ over an ensemble of the network {coupling} matrices for different values of the average degree. As reported in \cite{balance_cond}, a high value of $S_{\mathrm max}$ leads to the Fr\'echet distribution. $S_{\mathrm max}$ has the largest value for $\langle k \rangle=6$, as shown in Fig.\ref{cmax}. The value of $S_{\mathrm max}$ decreases for the lower average degrees displaying Gumbel (for $\langle k \rangle=4$) and Weibull (for $\langle k \rangle=2$) statistics. We further observe an interesting convergence behaviour of the shape parameter $\xi$ characterizing the GEV distribution for $R_{\mathrm max}$ as we increase the size of the network, keeping average degree of the network fixed. We address the size impact on $R_{\mathrm max}$ distribution in a separate section.

Next, we compare the extreme value statistics of the isolated SF network with that of the ER networks. For
$p_{in}=0.5$, the ER networks exhibit the Weibull statistics for $\langle k \rangle=6$ and show a transition 
to the Fr\'echet via the Gumbel distribution as the average degrees increases \cite{extrm_val}.
Again, the behaviour of $S_{\mathrm max}$ can be used to explain appearance of the Weibull (for ER networks) and 
the Fr\'echet (for SF network) distribution at $p_{in}=0.5$ for {a} fixed network size and the average degree.
Due to the hub-like structure, SF networks have a much higher largest degree than that of the corresponding 
ER networks. The high degree nodes of SF network ensures that $S_{\mathrm max}$ for {an} ER 
network will be much lesser than that of the SF network with the same average degree [Fig.\ref{cmax} (b)]. 
The high $S_{\mathrm max}$ values shift the shape parameter towards a more positive value yielding the
Fr\'echet distribution \cite{balance_cond}.

\begin{largetable}
\caption{Estimated parameters of KS test for fitting GEV and normal distributions of $R_{\mathrm max}$ for different
inhibitory inclusion probability ($p_{\mathrm in}$) of SF network over 5000 population. Other parameters are network size $N = 100$ and average degree $\langle k\rangle$ = 4.} 
\begin{tabular} { c  c  c  c  c  c  c  c } 
\hline  \specialrule{0.5pt}{0pt}{0pt}
\scriptsize $p_{\mathrm in}$& \scriptsize $\xi$ of GEV & \scriptsize $\sigma$ of GEV & \scriptsize $\mu$ of GEV  & \scriptsize p-value of KS test for GEV & \scriptsize $\mu$ of Normal & \scriptsize $\sigma$ of Normal & \scriptsize p-value of KS test for Normal\\ \hline \specialrule{1pt}{0pt}{0pt}

\scriptsize 0.0&		\scriptsize -0.2219&		\scriptsize 0.2109&		\scriptsize 7.302&		\scriptsize 0.1519&		\scriptsize 7.3843&		\scriptsize 0.2151&		\scriptsize 0.7846	\\  	
\scriptsize 0.2&		\scriptsize -0.3038&		\scriptsize 0.9238&		\scriptsize 4.8985&		\scriptsize 0.0161&		\scriptsize 5.2143&		\scriptsize 0.909&		\scriptsize 0.0005	\\  
\scriptsize 0.4&		\scriptsize -0.0071&		\scriptsize 0.6743&		\scriptsize 2.8222&		\scriptsize 0.0002&		\scriptsize 3.211&		\scriptsize 0.8428&		\scriptsize 0.0000	\\ 
\scriptsize 0.46&		\scriptsize 0.0617&			\scriptsize 0.4901&		\scriptsize 2.4659&		\scriptsize 0.3708&		\scriptsize 2.7812&		\scriptsize 0.6710&		\scriptsize 0.0000	\\ 
\scriptsize 0.48&		\scriptsize 0.0309&			\scriptsize 0.4590&		\scriptsize 2.4275&		\scriptsize 0.5492&		\scriptsize 2.7077&		\scriptsize 0.6074&		\scriptsize 0.0000	\\ 	
\scriptsize 0.50&		\scriptsize -0.0049&		\scriptsize 0.4333&		\scriptsize 2.4219&		\scriptsize 0.8788&		\scriptsize 2.6700&		\scriptsize 0.5521&		\scriptsize 0.0000	\\ 	
\hline \specialrule{0.5pt}{0pt}{0pt}
\end{tabular}
\begin{flushleft} 
\label{table.2} 
\end{flushleft}
\end{largetable}

\begin{figure}
 \centerline{\includegraphics[width=3.5in, height=1.2in]{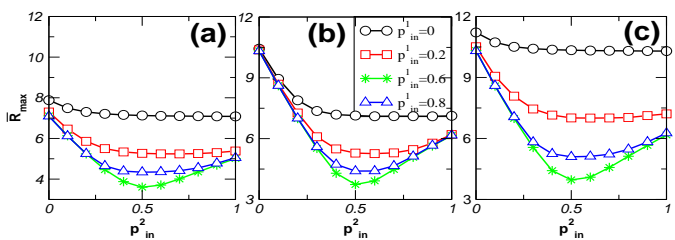}}
 \caption{(Color online) Average ${R}_{\mathrm max}$ as a function of inhibition probability ($P^2_{in}$). (a)ER-ER network of $\langle k \rangle$ = 6,
(b)ER-SF, and (c)SF-SF network where ($\overline{R}_{\mathrm max}$) for various inhibition probability for layer 1, $P^1_{in}$= 0($\circ$), 0.2($\square$),
0.6($\ast$), 0.8($\triangle$). Size of the network $N=100$.}
 \label{r_max_avgm}
\end{figure}
\begin{figure}
\centerline{\includegraphics[width=2.2in, height=1.3in]{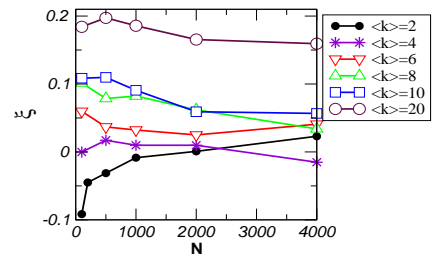}}
\caption{(Color online) Shape parameter of GEV distribution as a function of network size ($N$) for SF networks with average degree $\langle k \rangle =2 (\bullet)$, $\langle k \rangle =4 (\ast)$, $\langle k \rangle =6 (\triangledown)$, $\langle k \rangle =8 (\triangle)$,$\langle k \rangle =10 (\square)$, $\langle k \rangle =20 (\circ)$ for $p_{in}=0.5$.}
\label{size_effect}
\end{figure}
{\it Impact of Network Size on GEV statistics:}
Our investigations demonstrate a profound impact of directionality on $R_{\mathrm max}$ distribution as the network size is increased.
For $p_{in}=0$, i.e. without any inhibition in the network {coupling} matrix, the shape parameter exhibits a very small variation in its value as a function of network size. The distribution can be modeled using the Weibull statistics even for the large network size irrespective of the average degree of the network. As inclusion of the inhibitory coupling resulting in negative entries in the network {coupling} matrices introduces a change in the behaviour of the shape parameter. For $p_{in}=0.5$, there is an increment in the shape parameter as compared to that of $p_{in}=0$ for different average degrees and for a fixed network size. 
Moreover, we find that the shape parameter converges towards zero yielding the Gumbel distribution. For $N=100$, the shape parameter $\xi$ for the SF network displays Weibull (For $\langle k \rangle=2$), Gumbel (For $\langle k \rangle=4$) and Fr\'echet  (for $\langle k \rangle=6,8,10,20$) distribution. 
As we increase the network size, the value of shape parameter decreases for denser networks and takes a value which shows either the Gumbel distribution ((For example, $\xi=0.025$ for $\langle k \rangle=6$)) or a close resemblance with the Gumbel distribution (For example, $\xi=0.059$ for $\langle k \rangle=10$) {(Fig.~\ref{size_effect})}. Further, for sparser networks, there is an increment in the shape parameter as a function of network size leading to transition from the Weibull to the Gumbel distribution for larger networks. We find that irrespective of the nature of $R_{\mathrm max}$ distribution portrayed initially for small networks, increment in the network size leads to fluctuations in $R_{\mathrm max}$ at same scale. These results can be interpreted in terms of the stability of a system. Tail behaviour of the Fr\'echet and the Gumbel distributions being power law and exponential decay, respectively indicate that for the Gumbel distribution higher values of $R_{\mathrm max}$ is less probable. Hence large systems displaying Gumbel distribution is more stable than the corresponding systems with smaller network size which shows Fr\'echet statistics. Introduction of negative entries in network {coupling} matrices leads to a more stable system for the larger size and for
the denser networks.  

{\it Transition to Gumbel and Fr\'echet for multiplex network:} Next, we turn our attention to investigate 
the impact of multiplexing on $R_{\mathrm max}$ behaviour. It turns out that in absence of any inhibition, 
multiplexing {of a network with another network} leads to an enhancement in the $R_{\mathrm max}$ depending upon the architecture of the network it is multiplexing with and the impact of multiplexing is governed by the layer having the largest degree. A multiplexing {of ER network} with the SF network {leads to an}
enhancement in $R_{\mathrm max}$ value of the entire network
as compared to the isolated network, {with} the value of $R_{\mathrm max}$ lying approximately close to 
that of the isolated SF network as multiplexing increases degree of all the nodes in SF network by one only and 
hence, there is no profound impact on $R_{\mathrm max}$ {(Fig.~\ref{r_max_avgm})}.
As inhibitory nodes are introduced with probability $p_{in}$, there is a trade off between the inhibition and the multiplexing, resulting in a quite interesting behaviour of $R_{\mathrm max}$. Since, inhibition leads to make a matrix more anti-symmetric, the contribution from the layer having inhibition keeps on reducing as inhibition in that layer increases towards $p_{in}=0.5$ and impact of the layer having only symmetric coupling primarily governs $R_{\mathrm max}$.

When ER network is multiplexed with another ER network, inhibition in only one layer does not have any potential impact on $R_{\mathrm max}$ as it is governed by the another ER layer where inhibition {is not present. However,} for the ER-SF multiplex networks, behaviour of $R_{\mathrm max}$ strongly depends on the layer in which inhibition is introduced. For instance, if inhibition is introduced in the ER layer, $R_{\mathrm max}$ remains un-affected as it is governed by the SF layer and since no inhibition is present in SF, leading to no visible impact on $R_{\mathrm max}$. Value of $R_{\mathrm max}$ of the entire network, at no inhibition, is high due to this SF layer only. However, if inhibition is introduced in the SF layer, which leads to anti-symetricity in SF layer, that results in the decrease in the value of $R_{\mathrm max}$ of the entire network, finally reaching to a stable state when zero contribution comes from the SF layer (at $p_{in}=0.5$) and $R_{\mathrm max}$ of the entire network settles down to $R_{\mathrm max}$ of the isolated ER network. Thereafter, increasing inhibition in the SF layer does not have much impact as contribution to $R_{\mathrm max}$ comes from the ER layer as $R_{\mathrm max}$ of the isolated network for $p_{in}>0.5$ is always lower than that of the isolated ER network for $p_{in}=0$ {(Figs.~\ref{r_max_avg} and \ref{r_max_avgm})}.  

When both the layers are represented by SF networks, change in the $R_{\mathrm max}$ follows the similar behaviour as described for the ER-ER networks, as both the layers contribute equally to $R_{\mathrm max}$ and an inhibition in one layer makes another layer governing the $R_{\mathrm max}$ behaviour.

So far, we have discussed inhibition in only one layer of the multiplex network. As soon as we introduce inhibition in both the layers, $R_{\mathrm max}$ starts exhibiting several interesting phenomena which depends on trade-off of not only inhibition and multiplexing but trade-off between the inhibition in both the layers as well. Until inhibition in one layer (say 2), which has the larger highest degree, is lower than the inhibition in other layer, $R_{\mathrm max}$ keeps getting governed by the former and as soon as $p_{in}$ in layer 2 gets larger than $p_{in}$ of the layer 1, $R_{\mathrm max}$ starts getting governed by the layer 1. Interestingly, {minima} in $R_{\mathrm max}$ value, which a multiplex network can attain, is decided by the average degree of the multiplex network. The minima is reached, when inhibition in the SF layer becomes 0.5, as for this value maximum anti-symetricity is possible and consequently at this point $R_{\mathrm max}$ attains the minimum value governed by the average degree.

\begin{figure}
\centerline{\includegraphics[width=3.2in, height=1.4in]{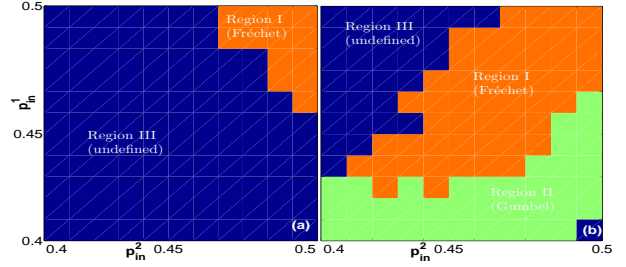}}
\caption{(Color online) Phase diagram depicting shape parameter $\xi$ for accepted GEV distribution {in $p_{in}^{(1)}-p_{in}^{(2)}$ plane} for (a) SF-SF and (b) ER-SF multiplex network. Regions I and II correspond to the Fr\'echet and the Gumbel distribution, respectively. Region III stands for undefined distributions. {$N=100$ in each layer and
average is taken over 5000 random realizations of the network.}}
\label{mul_phase}
\end{figure}

Next, we investigate fluctuations in the $R_{\mathrm max}$ behaviour for
multiplex networks having inhibitory nodes in both or one of the layers.
We present results for several possible 
combinations (ER-ER, ER-SF, SF-SF) and show that multiplexing with the similar type of the network does not lead to any significant impact on the $R_{\mathrm max}$ fluctuations, whereas multiplexing with a layer having a different network architecture has the following impact. The $R_{\mathrm max}$ statistics is again led by the layer having the largest degree as also found for the average $R_{\mathrm max}$ behaviour. But what is surprising that the range of $p_{in}$ for which the fluctuations can be modeled by the GEV statistics gets enhanced due to the multiplexing.
For example, if both the layers of the multiplex network are represented by the ER networks, the $R_{\mathrm max}$ statistics can be characterized by the Weibull distribution for {$p_{in}^{(1)}=0.5$} and {$0.4\le p_{in}^{(2)} \le 0.5$} \cite{SM_fig} as also {reflected} for the isolated case \cite{extrm_val}. If both the layers are represented by the SF networks, $R_{\mathrm max}$ of an ensemble of the multiplex network does not fit with any distribution (Region III, Fig.\ref{mul_phase}(b)) except at a narrow range around {$p_{in}^{(1)}=p_{in}^{(2)}=0.5$} (Region I, Fig.\ref{mul_phase}(a)) where it shows GEV statistics. Type of the statistics in this region depends upon the average degree of the network. As average degree increases, one gets a transition from the Weibull to the Fr\'echet distribution.

While $R_{\mathrm max}$ of the multiplex network having layers being represented with a similar architecture does not manifest any change as compared to that of the isolated case, an interesting phenomenon is observed for multiplex networks with different network topologies representing different layers. {An} ER network multiplexed with a SF network results in an increment in the shape parameter. Note that, the isolated ER and isolated SF exhibits the Weibull and the Fr\'echet statistics respectively. The multiplex ER-SF network exhibits a very different $R_{\mathrm max}$ fluctuation behaviour due to the interplay between the multiplexcity and the inhibition. For instance, Fig.\ref{mul_phase} depicts $R_{\mathrm max}$ fluctuation statistics of ER-SF network for average degree $\langle k \rangle=6$.  The $R_{\mathrm max}$ distribution can be characterized by the Gumbel (Region II, Fig.\ref{mul_phase}(b)) as well as the Fr\'echet (Region I, Fig.\ref{mul_phase}(b)) distribution in the range {$0.4\le p_{in}^{(1)}=p_{in}^{(2)}\le0.5$}. There is a transition from the Weibull to the Gumbel statistics for the parameter range {$0.4 \le p_{in}^{(1)} \le 0.5,p_{in}^{(2)}=0.42$}, arising due to the interplay between ER and SF layers. Because of the large degree nodes in the SF networks, we observe the Fr\'echet region for parameter range {$0.41 \le p_{in}^{(1)} \le 0.5$ and $0.42 \le p_{in}^{(2)} \le 0.5$}. Although, SF layer governs the $R_{\mathrm max}$ fluctuation 
behaviour, we find the Gumbel regime for a parameter range which reflects that the incremental behaviour of the shape parameter is greatly enhanced due to the multiplexing of a network with {a} different network architecture. This observation further indicates a reduction in the stability of a network as well as change in the dynamical behaviour of a network when it is multiplexed with another network having a different topology.

{\it Conclusion:} To summarize, we have investigated impact of inhibitory and excitatory coupling on $R_{\mathrm max}$ statistics for an ensemble of isolated and multiplex networks. 
While, the isolated SF networks for high connection density exhibit a transition from the Weibull to Fr\'echet distribution as a function of inhibition probability $p_{in}$, the SF networks with lower connection density does not show such transition for denser as well as sparser networks.  
In terms of stability of the corresponding network, this indicates that a higher values of $R_{\mathrm max}$ indicating a higher probability to became un-stable for denser network with fixed network size. Moreover, multiplexity is shown to have a strong influence on the stability of a network. The $R_{\mathrm max}$ behaviour deviates drastically when it is multiplexed with a network of different network topology. Our study demonstrates that the multiplex network becomes more unstable than the isolated network even when directionality was introduced in a single layer of the multiplex network.

Furthermore, we demonstrate a surprising effect of the network size on the $R_{\mathrm max}$ behaviour of the network at $p_{in}=0.5$. For denser networks, the $R_{\mathrm max}$ distribution is shown to have a transition from the Fr\'echet to the Gumbel distribution for larger networks. Whereas for sparser networks $R_{\mathrm max}$ exhibits a transition from Weibull to Gumbel as a function of network size indicating lesser stability. This indicates that the stability of the system increases for large dense network where as it decreases for large sparser networks. Fluctuations in the largest eigenvalue converges to a particular distribution for larger networks irrespective of the connection density.

The extreme value statistics tools has found its applicability in a wide range of real world systems. 
The largest mass distribution in mass transport network exhibits the Weibull, Gumbel or Fr\'echet statistics depending on the critical mass \cite{trans1}. Further, the level density of non-interacting bosons is shown to follow GEV distributions \cite{trans2}. Our work extends the application of extreme value statistics to the multiplex networks. 
The investigation presented here can be used to understand stability and dynamical behaviour of complex systems
having multiplex architecture and can be extended further to understand {effect of different layers 
interacting with inhibitory connections} on collective properties of real world systems having inherent multilayer architecture.

{\it Acknowledgment:}
SJ thanks DST project grant (EMR/2014/000368) for financial support and useful discussions with the group of 
Murilo Baptista during a visit to University of Aberdeen. SG and SKD acknowledge, respectively, DST Gov. of India for INSPIRE fellowship (IF150149) and Uni. Grants Commission, India.



\begin{thebibliography}{99}

\bibitem{barabasi_review} \Name{Albert R. \and Barab\'asi A.-L.} {\it Rev. Mod. Phys.} {\bf 74} (2002) 47-97 .

\bibitem{sync_osc}  \Name{Restrepo J. G.} {\it et al.,} {\it Phys. Rev. E} {\bf 71} (2005) 036151.

\bibitem{neural} \Name{Sompolinsky H.} {\it et al.,} {\it Phys. Rev. Lett.} {\bf 61} (1988) 259; \Name{Hennequin G.} {\it et al.,} {\it Phys. Rev. E} {\bf 86} (2012) 011909.


\bibitem{may_stability} \Name{May R. M.} {\it Nature} {\bf 238} (1972) 413-4.


\bibitem{mul_org_1} \Name{Boccaletti S.} {\it et al.,} {\it Phys. Rep.} {\bf 544} (2014) 1-122.

\bibitem{delay_multiplex} \Name{Singh A.} {\it et al.,} {\it EPL} {\bf 111} (2015) 30010 and reference therein.




\bibitem{mul_exp}\Name{G$\acute{o}$mez-Garde$\tilde{n}$es J.} {\it et al.,} {\it Sci. Rep.} {\bf 2} (2012) 620; \Name{Granell C.} {\it et. al.,} {\it Phys. Rev. Lett.} {\bf 111} (2013) 128701; \Name{Aguirre J.} {\it et al.,} {\it Phys. Rev. Lett.} {\bf 112} (2014) 248701.

\bibitem{inhibition} \Name{Compte A.} {\it et al.,} {\it J. Neurophysiol.} {\bf 89} (2003) 2707; \Name{Pilosof S.} {\it et. al.,} {\it arXiv} (2015) 1511.04453; \Name{Dwivedi S.} \and \Name{Jalan S.} {\it arXiv} (2016) 1604.07603.

\bibitem{gev_book}E. J. Gumbel, {\it Statistics of Extremes}
(Echo Point Books and Media, 2013).
 

 


\bibitem{config} \Name{Molloy M. \and Reed B.} {\it Random Struct. Algorithms} {\bf 6} (1995) 161-80.

\bibitem{pin_prob} \Name{Rajan K.} \and \Name{Abbott L.F.} {\it Phys. Rev. Lett.} {\bf 97} (2006) 188104; 
\Name{Strata P.} \and \Name{Harvey R.} {\it Brain Res Bull} {\bf 50 (5-6)} (1999) 349–50. 

\bibitem{cou_mat_PRE} \Name{Jalan S.} {\it et. al.,} {\it Phys. Rev. E} {\bf 84} (2011) 046107.
 

\bibitem{KS_test} Using KS test, we  fit the distribution with GEV statistics using functions kstest, gevfit and gevpdf from MATLAB statistics toolbox. Calculation of GEV distribution parameters are done with 95\% confidence level. 



\bibitem{GEV_normal} \Name{F$\ddot{u}$redi Z.} \and \Name{Koml\'os J.} {\it Combinatorica} {\bf 1} (1981) 233; \Name{Krivelevich M.} \and \Name{Sudakov B.} {\it Comb. Probab. Comput.} {\bf 12} (2003) 61.

\bibitem{edge_behaviour} \Name{Farkas I.J.} {\it et.al.,} {\it Phys. Rev. E} {\bf 64} (2001) 026704.

\bibitem{balance_cond} \Name{Jalan S. \and Dwivedi S.} {\it Phys. Rev. E.} {\bf 89} (2014) 062718.

\bibitem{extrm_val} \Name{Dwivedi S. K. \and Jalan S.} {\it Phys. Rev. E.} {\bf 87} (2013) 042714.

\bibitem{SM_fig} \href{https://figshare.com/s/66ef014328dcf2e9c501}{Supplementary Material}.

\bibitem{trans1} \Name{Evans M. R. \and Majumdar S. N.} {\it J. Stat. Mech. Theory Exp.} {\bf 2008} (2008) P05004.

\bibitem{trans2} \Name{Comtet A.} {\it et al.,} {\it Phys. Rev. Lett.} {\bf 98} (2007) 070404.

\end{thebibliography}
\end{document}


\begin{centering}
{\bf {Supplementary Material}}\\
\end{centering}
\begin{flushleft}
{\large 
\textbf{Interplay of inhibition and multiplexing : largest eigenvalue statistics}
}
\\
Saptarshi Ghosh,
Sanjiv K. Dwivedi,
Mikhail V. Ivanchenko,
Sarika Jalan$^{\ast}$

\vspace{0.4cm}

$\ast$ E-mail: sarikajalan9@gmail.com
\end{flushleft}

\begin{figure}[h]
\centerline{\includegraphics[width=3.8in, height=3.2in]{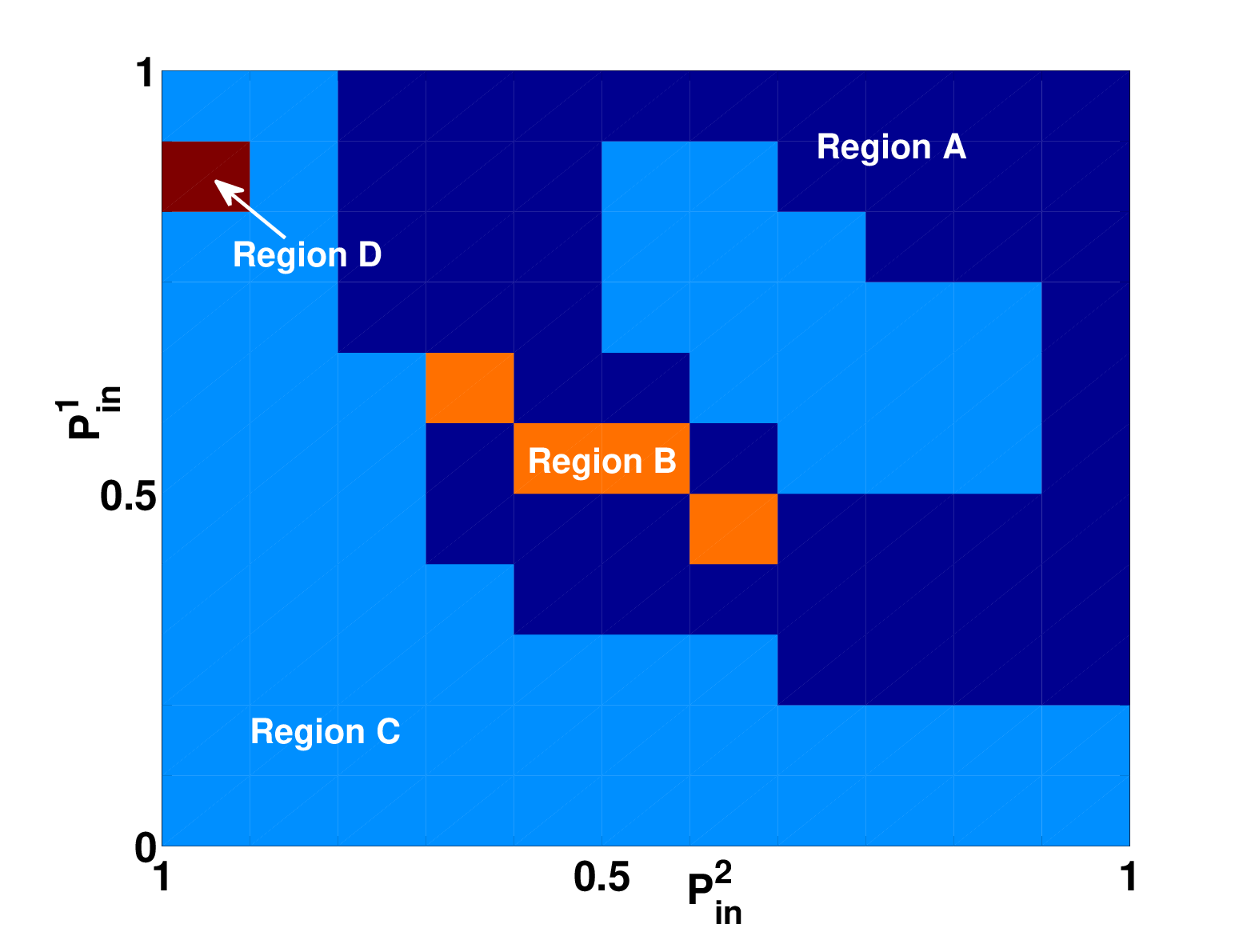}}
\caption{(color online) Phase diagram depicting shape parameter $\xi$ for accepted GEV distribution for ER-ER multiplex network as a function of IC inclusion probabilities ($p_{in}$) in both the layers. Region B corresponds to the Weibull. Region A stands for undefined distributions. Size of the network N=100 in each layer.}
\label{mul_phase_er_er}
\end{figure}

\begin{figure}[h]
	\centerline{\includegraphics[width=3.8in, height=3.2in]{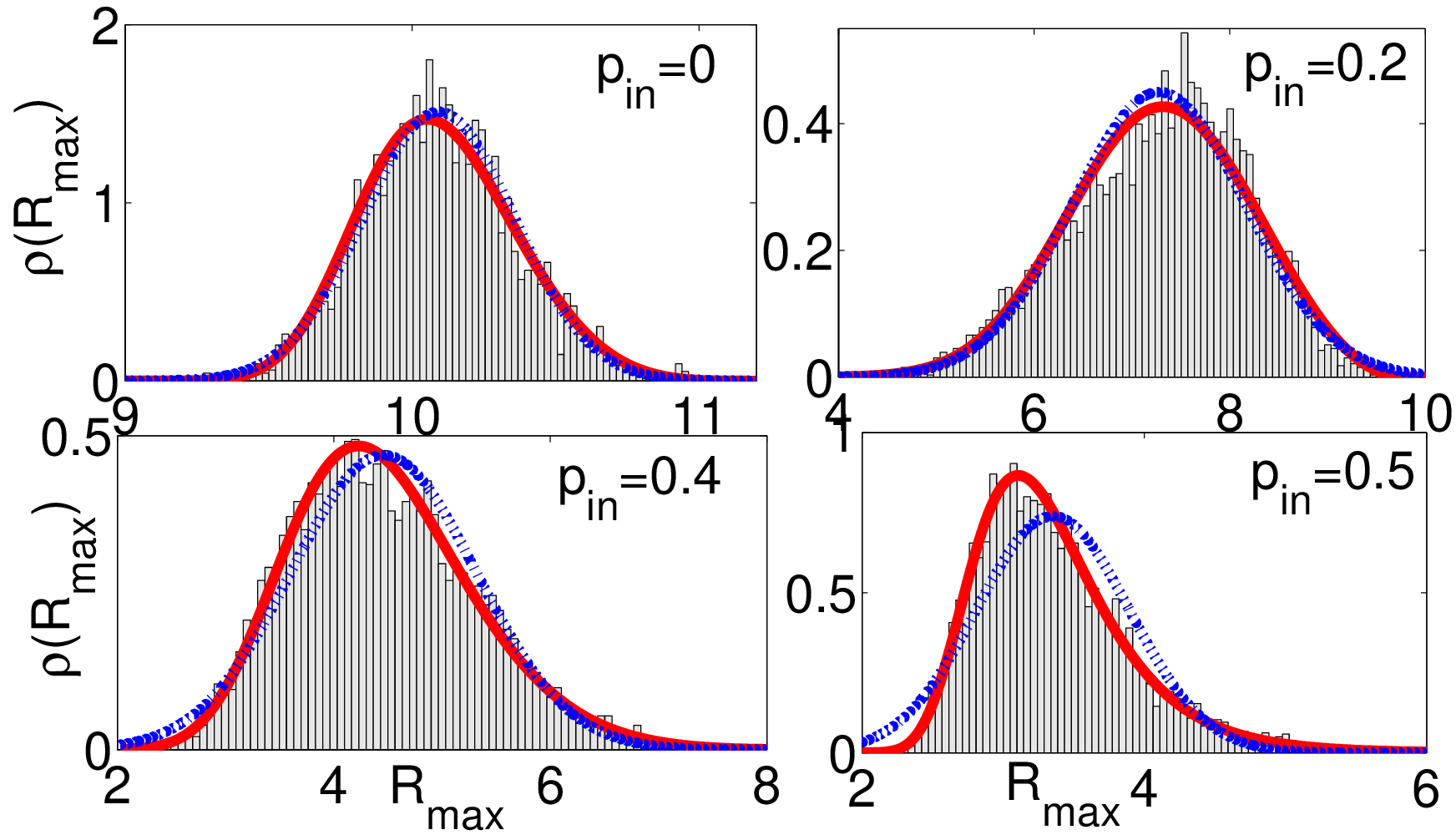}}
	\caption{(Color online) Distribution of $R_{max}$ of SF networks with average degree $\langle k\rangle=4$ for various IC inclusion probabilities ($p_{in}$). 
		Histogram is fitted with normal (blue dotted line) and GEV (red solid line) distributions. Network size N=500.}
	\label{hist_n500_k4}
\end{figure}

\begin{figure}[h]
	\centerline{\includegraphics[width=3.8in, height=3.2in]{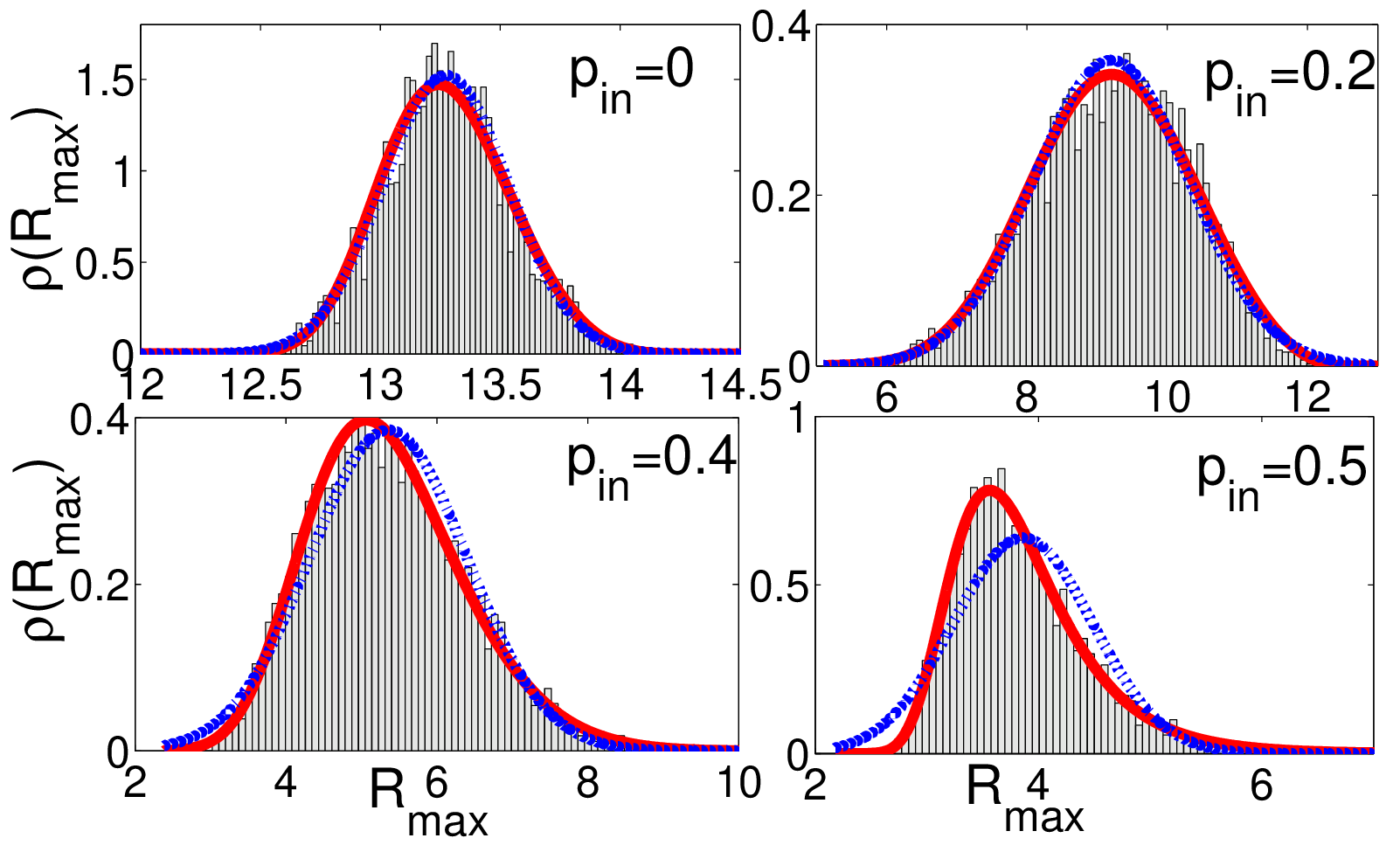}}
	\caption{(Color online) Distribution of $R_{max}$ of SF networks with average degree $\langle k\rangle=6$ for various IC inclusion probabilities ($p_{in}$). 
		Histogram is fitted with normal (blue dotted line) and GEV (red solid line) distributions. Network size N=500.}
	\label{hist_n500_k6}
\end{figure}

\begin{table}[ht]
	\caption{Estimated parameters of KS test for fitting GEV and normal distributions of $R_{\mathrm max}$ for different network sizes of SF network over a average of 5000 random realization. Other parameters are inhibition inclusion probability $p_{in} = 0.5$ and average degree $\langle k\rangle$ = 6.}
	\begin{tabular} { p{1cm} p{1.6cm}  p{1.4cm}  p{1.4cm}  p{2.7cm}  p{1.8cm}  p{1.8cm}  p{2.7cm} } 
		\hline 
		$N$&    $\xi$ of GEV &     $\sigma$ of GEV &   $\mu$ of GEV&   p-value of KS test for GEV &    $\mu$ of Normal&    $\sigma$ of Normal&   p-value of KS test for Normal\\ \hline 
		100&	0.0598&	       0.4807&	       2.7906&	   0.9688&	                   3.0983&	       0.6685&	         0.0000  \\ 
		500&	0.0367&	       0.4727&	       3.5745&	   0.9223&	                   3.8655&	       0.6349& 	         0.0000  \\
		1000&	0.0323&	       0.4737&	       3.8953&	   0.4531&	                   4.1849&	       0.6303&	         0.0000  \\
		2000&	0.0247&	       0.4671&	       4.1003&	   0.192&	                   4.3825&	       0.6126&	         0.0000  \\
		4000&	0.0409&	       0.5164&	       4.5638&	   0.2938&	                   4.8846&	       0.6893&	         0.0000  \\
		\hline 
	\end{tabular}
	\begin{flushleft} 
		\label{table.1} 
	\end{flushleft}
\end{table}